# Metallic coplanar resonators optimized for low-temperature measurements

Mojtaba Javaheri Rahim[1], Thomas Lehleiter[1], Daniel Bothner[2], Cornelius Krellner[3], Dieter Koelle[2], Reinhold Kleiner[2], Martin Dressel[1] and Marc Scheffler[1]

[1]1. Physikalisches Institut, Universität Stuttgart, Pfaffenwaldring 57, 70569 Stuttgart, Germany
[2]Physikalisches Institut and Center for Quantum Science in LISA[+], Universität Tübingen, 72076 Tübingen, Germany
[3]Institute of Physics, Goethe University Frankfurt, Max-von-Laue-Strasse 1, 60438 Frankfurt am Main, Germany

**Abstract**

Metallic coplanar microwave resonators are widely employed at room temperature, but their low-temperature performance has received little attention so far. We characterize compact copper coplanar resonators with multiple modes from 2.5 to 20 GHz at temperatures as low as 5 K. We investigate the influence of center conductor width (20 to 100 μm) and coupling gap size (10 to 50 μm), and we observe a strong increase of quality factor ($Q$) for wider center conductors, reaching values up to 470. The magnetic-field dependence of the resonators is weak, with a maximum change in $Q$ of 3.5% for an applied field of 7 T. This makes these metallic resonators well suitable for magnetic resonance studies, as we document with electron spin resonance (ESR) measurements at multiple resonance frequencies.

## 1. Introduction

Coplanar waveguides (CPWs) from superconducting [1, 2] and metallic (non-superconducting) materials [3-5] are widely used in microwave transmission lines. Since conductor strip and ground planes are on the same side of the substrate, they can be easily adapted to active/passive components. For these reasons CPWs are suitable for classical engineering applications such as sensors [6] and filters [7] as well as new modern physics experiments like cold atoms current traps [8] and quantum information processing [9]. The electric and magnetic energy of the carrying signal can be efficiently coupled to the surrounding medium, and by proper design these fields can have the desired orientation. These characteristics are prerequisites for, e.g. electron spin resonance (ESR) experiments [10]. CPW resonators can support multiple resonances, and CPW devices with miniaturized size can be employed in the mK temperature range experiments using dilution refrigerators [9, 11-15]. These points become important when the electronic and magnetic properties of a material have to be characterized as a function of temperature, frequency and magnetic field. There are no reports explicitly exploring the performance of metallic CPW resonators at temperatures lower than liquid nitrogen temperatures [5]. We have optimized the parameters of metallic CPW microwave resonators for low-temperature use and present the results in this paper.

## 2. Experimental

In the functionality of a typical planar resonator, the most decisive factors are coupling gaps [16, 17], center conductor width [18, 19], film quality and edge sharpness [20, 21], ground plane size and packaging effects [22], effective permittivity ($\varepsilon_{eff}$) [4] and meander line dispersion [23, 24]. Coplanar resonators can be designed based on these theoretical considerations, but resonator optimization for new applications often involves additional testing and tuning of design parameters of the actually fabricated devices, such as presented in this work. Since we are mainly interested in low-temperature experiments, our CPW resonators were characterized at temperatures down to 5 K. We designed, fabricated and measured several sets of resonators (characteristic impedance 50 Ω) from 2.5 to 20 GHz at temperatures down to 5 K. All resonators were designed as open-ended half wave (λ/2) resonators for their fundamental frequency (consequently, the "mode $n$" of the series of harmonic resonance frequencies is expected when the relationship $L = n\lambda/2$ is fulfilled where $L$ is the center conductor length), and they are capacitively coupled to input and output CPWs via two identical gaps. Figure 1(a)-(c) show schematically the structure and the geometrical parameters of the chips. The miniaturization of the resonator size is necessary with regard to the small volumes of the samples which are supposed to be investigated via the resonators. In order to reach high filling fraction of the sample volume with respect to the resonator mode volume, the center conductors of the resonators were meander-like folded as depicted in figure 1(a); with the length of $L$=2.55 cm, expecting a fundamental mode (first mode) of 2.5 GHz. In addition, the small size facilitates implementation in superconducting magnets and/or dilution refrigerators.

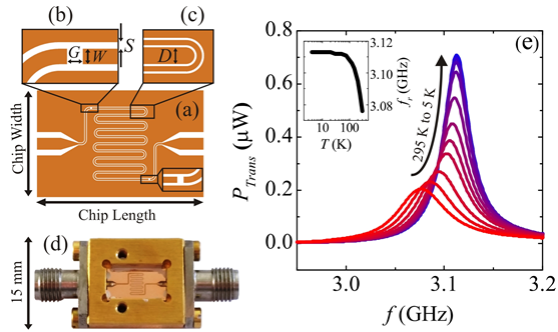

**Figure 1.** (a) The meandered structure of the fabricated CPW resonators. The orange parts show the metallization (copper). The geometrical parameters are shown in the enlarged sketches of (a), (b) and (c). $G$ stands for the coupling gaps, $W$ is the center conductor width, $S$ shows the distance between the center conductor and the ground planes, and $D$ is the width of the ground plane between two adjacent lines of the meander structure. The photograph (d) shows the brass box in which a CPW resonator chip is placed. The SMA connectors guide the microwave signal into and out of the resonator. (e) Temperature dependence of the fundamental mode of an electroplated resonator with $W$=60 μm and $G$=40 μm. (Inset) Resonance frequency vs. temperature.

Table I shows the indicated dimensions and distances in figures 1(a)-(c) for each set. The $D/(W + 2S)$ ratio is bigger than 1 for all the sets which is enough to neglect the ground plane width effect on the characteristic impedance [25]. The resonators were fabricated either from a sputtered copper film ("sputtered resonators") or by a copper electroplating process ("electroplated resonators"). The comparison between the sputtered and electroplated resonators provides information about the influence of the surface roughness, the edge quality of the center conductor (and the ground planes), and the residual resistivity by which we can reach higher quality factors at low temperatures. Low-loss polished sapphire ($\varepsilon_r$≈10.5 [26], thickness 430 μm) was used as substrate. For the sputtered resonators it was covered with 1 μm copper film (resistivity $\rho_{5K}$=0.0510 μΩcm, residual resistance ratio $RRR$≈29, skin depth $\delta_{5K,2.5GHz}$ ≈227 nm) with a 5 nm chromium buffer layer. Standard UV photolithography and Ar-ion beam etching processes were used to form the resonators. In the case of the electroplated resonators, the photolithography was first performed on a 50 nm evaporated copper film with 5 nm chromium buffer layer. Afterwards the structure was placed in an electroplating bath to grow a 1 μm copper film ($\rho_{5K}$=0.0415 μΩcm, $RRR$≈50, $\delta_{5K,2.5GHz}$≈205 nm). After removing the resist, Ar-ion beam etching removed the 50 nm copper film and the buffer layer between the center conductor and the ground planes.

**Table 1.** The geometrical parameters of the fabricated resonators according to figures 1(a)-(c): $W$ is the width of the center conductor, $S$ denotes the distance between the center conductor and the ground planes, $G$ indicates the gap size, $D$ is the inner diameter of a turn. "10:50" means 10, 20, 30, 40 and 50 μm.

| $W$ (μm) | $S$ (μm) | $G$ (μm) | $D$ (μm) | $\dfrac{D}{W+2S}$ | Chip Size (mm$^2$) |
|---|---|---|---|---|---|
| 20 | 8 | 10:50 | 264 | ≈7.33 | 3×4 |
| 40 | 16 | 10:50 | 226 | ≈3.14 | 3×4 |
| 60 | 24 | 10:50 | 190 | ≈1.76 | 3×4 |
| 80 | 32 | 10:50 | 240 | ≈1.67 | 6×10.5 |
| 100 | 40 | 10:50 | 300 | ≈1.67 | 6×10.5 |

Resonators of $W$=20 to 60 μm were fabricated all together on a single piece of sapphire; the same was done for the resonators of $W$=80 and 100 μm. The processed chips were placed in especially designed brass boxes shown in figure 1(d). Standard SMA connectors were connected to the chips in the box by H20E-FC EPO-TEK® conductive adhesive. The chip sizes (box sizes) were kept small to shift parasitic modes to higher frequencies (4×7 mm$^2$ for $W$=20 to 60 μm and 6×10.5 mm$^2$ for $W$=80 and 100 μm).

## 3. Results and Discussion

The microwave measurements were performed by utilizing an Agilent® microwave source and power meter. With a Janis® helium flow cryostat we reach temperatures down to 5 K. The loaded quality factors were calculated by $Q = f_r/\Delta f$ where $f_r$ is the resonance frequency and $\Delta f$ is the half power bandwidth of the resonance signal. All the *Q* values quoted throughout this article are loaded quality factors, which approximately equal the internal quality factors of our resonators.[1]

The fundamental frequency of the resonators slightly shifts up by increasing *W*+2*S* (from ~2.60 GHz for *W*=20 μm and *S*=8 μm to ~3.15 GHz for *W*=60 μm and *S*=24 μm). This is due to the decrease of effective permittivity ($\mathcal{E}_{eff}$) by increasing *W*+2*S* [3, 4, 22, 25]. The fundamental for both *W*=80 and 100 μm are ~3 GHz. The thickness in these cases is about 100 nm lower (compared to those with *W*=20 to 60 μm) which causes $\mathcal{E}_{eff}$ to increase and thus the fundamental mode is below ~3.15 GHz (of *W*=60 μm) [3, 4]. The temperature dependence of the resonance frequency is plotted in figure 1(e); the rightward shift of the resonance frequency upon cooling has several reasons: (i) thermal contraction, (ii) $\mathcal{E}_{eff}$ of CPW decreases in absolute value as the conductivity of the metallic conductor increases [3] and (iii) the relative permittivity of sapphire substrate decreases with lowering the temperature [26]. When the resonator is cooled down, the quality factor increases significantly: at liquid nitrogen temperatures *Q* is more than 2.5 times the room temperature value and it can reach an enhancement of almost a factor of 4 upon further cooling as shown in figures 2(a) and (b). This increase of the quality factor comes to a halt around 20 K because the residual-resistance regime, in which the resistivity of the copper film is governed by defects and does not decrease by further lowering the temperature [27], is reached.

Microwave losses in metals are supposed to grow with frequency; thus for the higher modes a lower quality factor is expected. However, for example for the resonators of *W*=60 μm, independent of the *G* values and fabrication method, the highest *Q* values are observed at the 3$^{rd}$ mode (~8.4 GHz); see figures 2(c) and (d). Such behavior is found in all five sets of resonators, i.e. mode 2 (~5 GHz; for *W* ≤ 40 μm) or mode 3 (~8 GHz; for *W* ≥ 60 μm) have the highest *Q* values. This is in contrast to basic considerations of metallic CPWs, where the losses should monotonously increase with increasing frequency, leading to a monotonous decrease of *Q*. The reason behind the observed non-monotonic behavior might be found in the details of the meander structure, but a thorough understanding might require additional studies. Similar observations have been reported previously on aluminum CPW resonators of much larger dimensions and at room temperature [30].

---
[1] The measured, loaded quality factor $Q_L$ is a combination of internal and external quality factors, $Q_{int}$ and $Q_{ext}$, respectively [28]: $\frac{1}{Q_L} = \frac{1}{Q_{int}} + \frac{1}{Q_{ext}}$. Considering the coupling capacitance of our resonators (which amounts to approximately 10 fF for the extremal case of *G*=10 μm for *W*=100 μm) [29], we expect $Q_{ext}$ well above 20000 for all studied resonances, which is much higher than the measured $Q_L$, and therefore we can neglected $Q_{ext}$ with respect to $Q_{int}$.

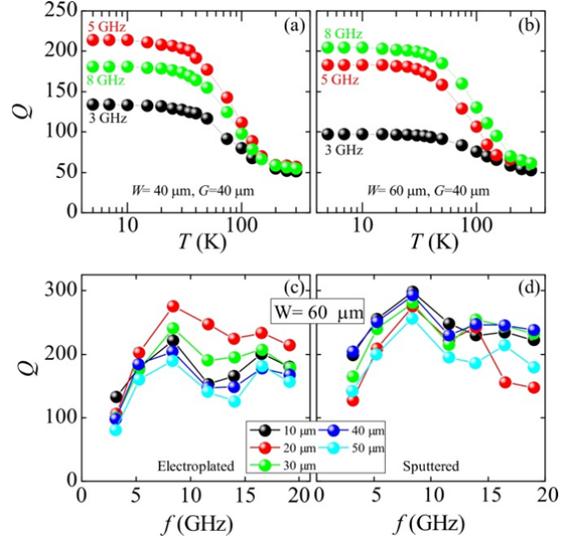

**Figure 2.** Temperature (a, b) and frequency (c, d) dependence of the quality factor *Q* for different resonator modes and structures. The first three modes of an electroplated resonator with *W*=40 μm and *G*=40 μm (a), and *W*=60 μm and *G*=40 μm (b) exhibit increasing *Q* upon cooling with saturation in the residual-resistance regime. The quality factor for different modes of (c) electroplated, and (d) sputtered resonators with *W*=60 at 5 K. The legend refers to the *G* values.

Figure 3 displays the quality factor values for the modes 2 and 3 of all the resonators. The values of between ~300 (*W*=60 μm and *G*=10 and 30 μm) and 470 (*W*=100 μm and *G*=20 and 40 μm) are the highest values observed for a metallic CPW resonator with narrow center conductor widths of 60 to 100 μm. We reach *Q*=120 at room temperature for a resonator with *W*=100 μm and *G*=20 and 50 μm. As far as the design parameters of the resonator are concerned, the center conductor width *W* clearly is the most relevant in obtaining high *Q* values; we did not observe a pronounced dependence on *G* or *D*/(*W* + 2*S*).

The sputtered resonators have slightly higher *Q* compared to the electroplated resonators, although the sputtered films have a higher residual resistivity. The higher *Q* values of the sputtered resonators can be explained by the lower surface roughness of the sputtered films compared to the electroplated films (13 nm vs. 60 nm from AFM) and could also be due to the better edge definition [21]. The sharper and straighter line edges in the sputtered resonators are displayed in the SEM images of figure 4. The redeposited thin raised walls (non-removable by ultrasonic vibration) during the ion milling process on the edges of the sputtered lines [31] makes it difficult to discuss the edge definition role in the achieved higher $Q$.

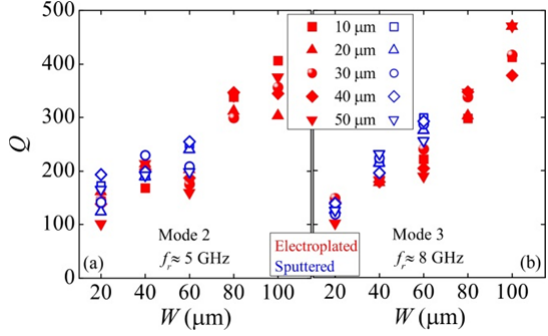

**Figure 3.** The enhancement of $Q$ by widening the center conductor plotted for mode 2 (a), and mode 3 (b) at $T$=5 K. The different symbols refer to different $G$ values and films as indicated.

Many physical experiments for which CPW structures are utilized, e.g. ESR studies, have to be performed in the presence of an external magnetic field. Therefore, we studied the influence of an external static magnetic field on our resonators.

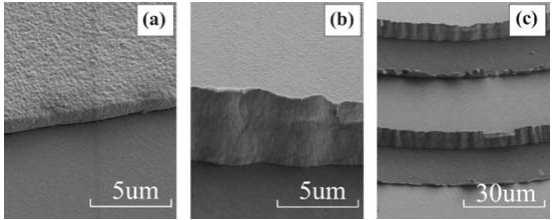

**Figure 4.** SEM images of (a) an electroplated line edge, and (b) sputtered line edge. Panel (c) shows the upraised edges on both the center conductor and the ground plane of a sputtered resonator.

Figure 5(a) shows the relative change of $Q$ in a magnetic field applied in the plane of the resonator, parallel to the meander lines [see the inset of figure 5(a)]. Even at fields as high as 7 T, the reduction in $Q$ is less than 3.5% for all the modes, in contrast to the massive suppression of $Q$ in magnetic fields typically found in superconducting resonators [28, 32-34]. The observation can be explained by the magnetoresistance of copper at higher fields [35]. Planar structures (microstrip, stripline and coplanar) have already been employed for ESR experiments [15, 36-46]. Although microstrip and stripline resonators have also been successfully applied in ESR measurements [36-38, 41, 45, 47], coplanar configurations might be more appropriate candidates since they carry the signal non-dispersively in contrast to a microstrip configuration [45, 48].

Superconducting planar resonators can be substantially limited by their critical fields [26, 35], and broadband transmission line structures [15, 44] have limitations in sensitivity. In order to test whether our metallic CPW resonators can be employed for ESR experiments, a 2,2-diphenyl-1-picrylhydrazyl (DPPH) sample ($200\times100$ μm$^2$) was placed on the middle of the resonator meander line and fixed by a small amount of vacuum grease. The obtained ESR lines are plotted in figure 5(b); here we plot the change in $Q$ (i.e. $Q_B$-$Q_{B=0}$) vs. relative magnetic field ($B$-$B_{Resonance}$) for various resonance fields (depending on frequency). We measured the resonator transmission at each magnetic field while keeping the temperature constant at 5 K in a cryostat equipped with a superconducting magnet. As expected, the higher intensity of the ESR signal is observed at the higher field/frequency. The extra peaks on the sides of the ESR signals are due to the narrower ESR line width than the resonator line width.

We have performed another series of the ESR experiment on the heavy fermion compound, YbRh$_2$Si$_2$ ($3 \times 4$ mm$^2$); a rare example where an ESR investigation of the Kondo ions' spin dynamics is feasible [49]. ESR studies have already been done on this material using conventional and superconducting planar resonators [45, 49]. In figure 5(c) we plot the ESR lines obtained at 1.7 K for two different resonance fields. The lines exhibit a Dysonian shape typical for metals. The deeper lines at higher resonance fields/frequencies demonstrate the higher intensity of the ESR signals as already anticipated.

## 4. Conclusion

In conclusion, we have characterized, improved, and utilized miniaturized metallic CPW resonators for cryogenic measurements. Significant improvement of the quality factors is realized by cooling down the resonators to liquid helium temperatures; we reached $Q$ values as high as 470 whereas the miniaturized size of the resonators limited the center conductor width not to be wider than 100 μm. Compared to the electroplated resonators, lower surface roughness and probable edge definition effects of the sputtered resonators could improve the quality factors up to 10%, although their film resistivity is slightly higher. We demonstrate the resonators working up to 7 T which is much further than the limits of similar conventional superconducting resonators (but reachable with cuprate devices) [34, 43, 46]. The values of the quality factor drop at 7 T by less than 3.5% which exert a very light background on the obtained spectra. This background is even much weaker (~0.4%) at 1 T. Note that for field above 1 T conventional superconducting resonators suffer either of an extreme loss of the quality factor or stopping any transmission. Our metallic CPW resonators were successfully employed in several low-temperature and low-frequency ESR experiments. If performed in a dilution refrigerator we expect even more interesting physics of exotic behavior of materials such as YbRh$_2$Si$_2$ [45, 50, 51].

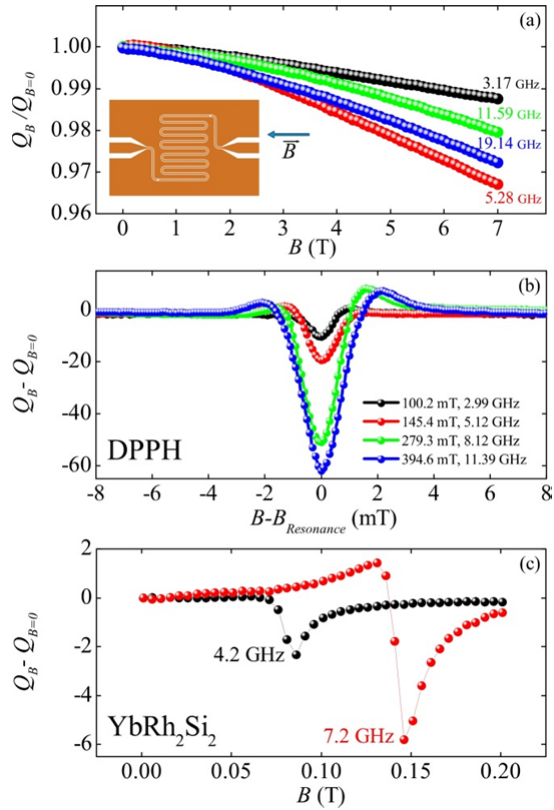

**Figure 5.** (a) Normalized $Q$ values decay at external magnetic field. The data are taken at $T$=5 K with the resonators of $W$=60 μm and $G$=20 μm. (b) ESR resonance curves of a DPPH sample at 5 K using a resonator with $W$=60 μm and $G$=20 μm. Each data set belongs to a given resonance field ($B_{Resonance}$) as indicated in the legend. (c) Two ESR resonance curves of YbRh$_2$Si$_2$ at different frequencies taken at $T$=1.7 K with a resonator with $W$=40 μm and $G$=30 μm.

## Acknowledgements

This work was supported by the Deutsche Forschungsgemeinschaft (DFG), including SFB/TRR 21, and by the EU-FP6-COST Action MP1201. The authors thank Gabriele Untereiner for technical support, Ali Tavassolizadeh for providing the sputtered films, Rainer Stöhr and Roman Kolesov for support with the electroplating, and Harald Giessen for access to the cleanroom facilities.